# Monte Carlo Studies of a Novel LiF Radiator for RICH Detectors

A. Efimov, M. Artuso, Min Gao, R. Mountain, F. Muheim, Y. Mukhin, S. Playfer, and S. Stone

*Dept. of Physics, Syracuse Univ., Syracuse, NY, 13244-1130*

We show that a multifaceted LiF radiator produces more Cherenkov light and has better resolution per photon than a flat radiator slab when used in a ring imaging Cherenkov counter. Such a system is being considered for the CLEO III upgrade.

## I. Introduction

Ring imaging Cherenkov detectors (RICH) are capable of providing excellent identification of charged particles. Several systems have been implemented in hadron beams and $e^+e^-$ collider experiments [1]. Many of these have used liquid or gaseous freon radiators and have used TMAE vapor as the photosensitive element [2]. TMAE introduces special problems. Its relatively low vapor pressure requires a rather thick conversion volume ($\approx$10 cm) or high temperatures. Also, it is very corrosive, so that special handling precautions must be taken and there is evidence that it harms wire chambers.

A triethylamine (TEA) methane mixture is known to have usable quantum efficiency in the wavelength range between 135-165 nm. Liquid Freon radiators are not transparent in this wavelength region so a crystal radiator must be used. A RICH system with a LiF radiator and photon detector consisting of $CH_4$ and TEA vapor has been successfully tested by the Fast-RICH group at CERN [3]. With a prototype detector employing fast VLSI electronics, an average of 10.4 photoelectrons were detected, for an incident track angle of 25$^o$ with respect to the radiator, with a resulting resolution per track of 4.2 mr. The angle of Cherenkov radiation emitted by a charged track passing through the LiF is given by

$$cos(\theta_C) = 1/(n \cdot \beta), \qquad (1)$$

where $\beta = v/c$.

We use as a benchmark the separation between pions and kaons at a momentum of 2.8 GeV/c, which is the upper limit of particle momentum from $B$ decays from the $\Upsilon(4S)$ resonance at a symmetric $e^+e^-$ collider. Since LiF has an index of 1.5 at 150 nm, which is the center of the useful wavelength range in this system, the $K/\pi$ separation at 2.8 GeV/c is 12.8 mr. We define separation in terms of the number of

standard deviations as
$$N_\sigma = \frac{\theta_C(K) - \theta_C(\pi)}{\frac{1}{2}[\sigma(K) + \sigma(\pi)]}, \qquad (2)$$
where $\sigma$ refers to the rms error on the track angle measurement. The CERN test results correspond to an $N_\sigma$ of 3. While a device built with this resolution would give respectable results, our goal is to design a device where $N_\sigma$ equals 4.

## II. Flat Radiator Configuration

The detector we envision for the CLEO III upgrade fits between the CsI electromagnetic calorimeter and a new drift chamber [4]. It is approximately cylindrically symmetric with the LiF radiators in the form of tiles ($\approx$16x16 cm$^2$) at an inner radius of 82 cm and a gap of 16 cm between the radiator and the entrance window of the wire proportional chamber. The length of the radiators is 236 cm, while the photon detectors are 250 cm long. The photon detector is similar to that used in the CERN tests, but differs because the pads are 7.5 x 7.5 mm$^2$, and the pulse height on each pad is measured.

A reasonable extrapolation of the Fast-RICH prototype results shows that the photoelectron yield can be increased by 43%. This results from several factors: increase in the size of the detector area (10%), the CERN prototype was only 50 cm wide, not sufficient to contain the full image; having the chamber voltage on the plateau (8%), only after the test was it discovered that the voltage was a bit too low; cleaner expansion volume gas (5%); thinner CaF$_2$ windows and strips (8%); and connecting up all of the electronics channels (5%). The quantum efficiency assumed is taken as that found in [3].

A system of flat 1 cm thick LiF radiators must have the angle of the incident charged track be larger than about 6$^o$ with respect to the normal in order to avoid total internal reflection of all the Cherenkov light. Thus in the center of a cylindrically symmetric detector the radiators must be tilted. An angle of about 20$^o$ is required to have adequate Cherenkov light. Even so, most of the Cherenkov light is lost.

The angular resolution per detected photon is comprised of several sources. The most important are the chromatic error, which results from the variation of the index of refraction with the wavelength, the emission point error, which results from the lack of knowledge about where the photon is emitted, and the position error in detecting the photon. The individual sources of error determined by using GEANT are shown as a function of the track dip angle $\theta$ in Fig. 1. All calculations in this paper are done using 2.8 GeV/c pions.

This system has about 13.5-14 mr resolution per detected photon independent of the track incident angle. This corresponds to a 3.7 mr resolution per track

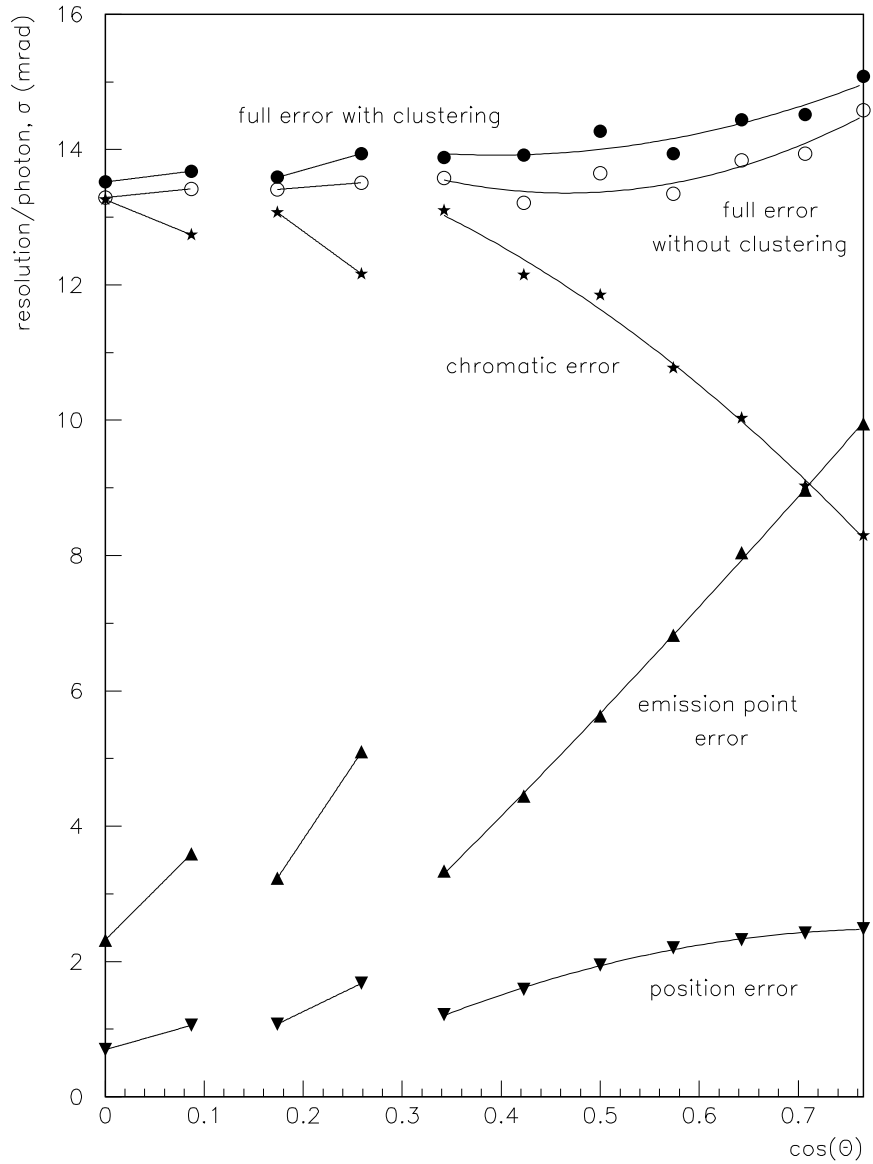

Figure 1: The individual sources of Cherenkov angle error per detected photon for a 10 mm thick flat LiF radiator. These include position determination error in the chamber, photon emission point error, chromatic error and overlap error due to some of the photons overlapping in the chamber. The breaks in the curves occur because the first two radiator sections are tilted at a 20° angle with respect to the incident track direction.

# III. "Sawtooth" Radiator Configuration

To get more light out of the LiF it is advantageous to facet the surface where the Cherenkov light exits. Two radiator designs with $45^0$ facets which we are considering are shown in Fig. 2. The first design has 5 mm deep facets, while the second has facets on the order of 1 mm or less in depth. The grooves run along the 236 cm length of the detector, i.e. along the $z$-axis. To explore the potential of such radiators, we performed Monte Carlo simulations of different facet angles always keeping the average thickness of the radiator at 10 mm. Although we have simulated both radiators, we show results only for the more deeply faceted one. The smaller facets give somewhat better performance in that the spread in thickness of the radiator is much smaller. Two quantities are of interest, the average number of photonelectrons and the angular resolution per photoelectron. The latter changes because of differences in the chromatic error, which is influenced by the angle of the photon with respect to the normal as it leaves the surface [5].

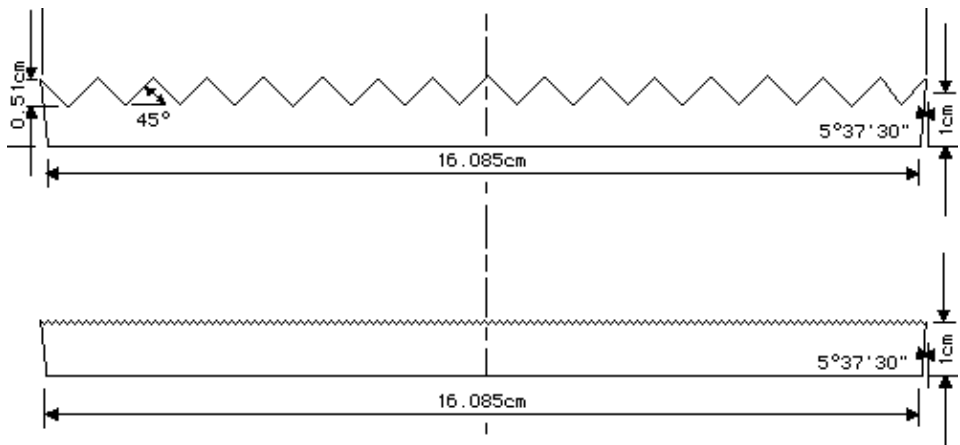

Figure 2: Two possible "sawtooth" designs. The bottom one has groves less than 1 mm in depth.

In order to compare different facet angles expeditiously, we did not use a full GEANT simulation, but only looked at the chromatic and emission point errors. In Fig. 3 we show the average number of photons leaving the surface, as a function of incident track angle, $\theta$, for different teeth angles, where larger angles refer to sharper teeth. Also shown is the flat radiator for the non-tilted sections. The optimum angle is close to $45°$. (Note, the Cherenkov angle is $48^0$ for relativistic tracks.) The vertical scale does not reflect the absolute number of photoelectrons. In Fig. 4 we show the resolution per photoelectron. Sharper tooth angles give better resolution. Combining these considerations, we find that the best performance in terms of resolution per track is given by $45°$ teeth.

We proceed by performing full GEANT level simulations on the $45°$ tooth angle radiator. The resolution per photoelectron, the number of photoelectrons and the Cherenkov angular resolution per track for $\theta$ equals $90°$ is shown on Fig. 5.

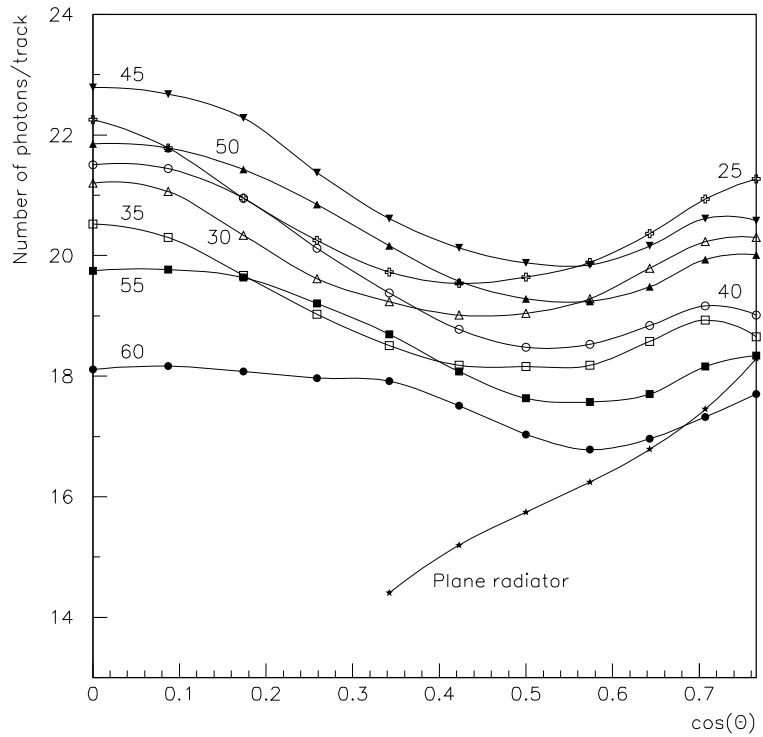

Figure 3: The average number of photons (relative scale) detected as a function incident track angle for different "tooth" angles.

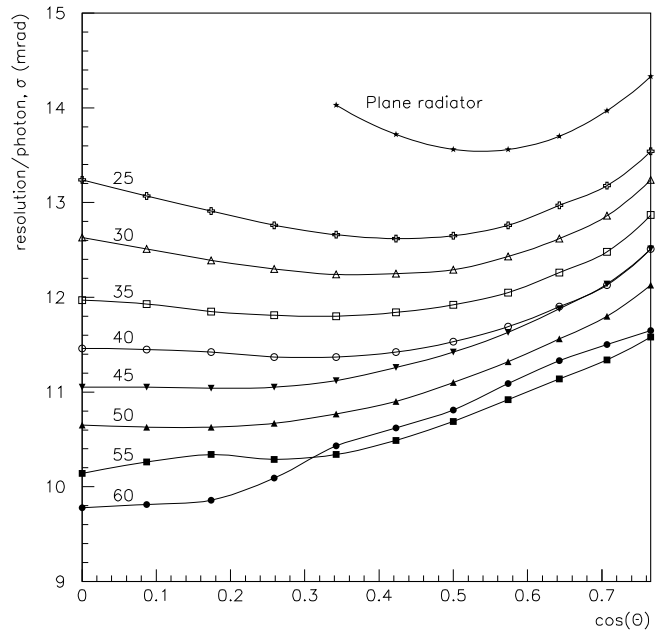

Figure 4: The angular resolution per photon as a function of incident track angle for different "tooth" angles.

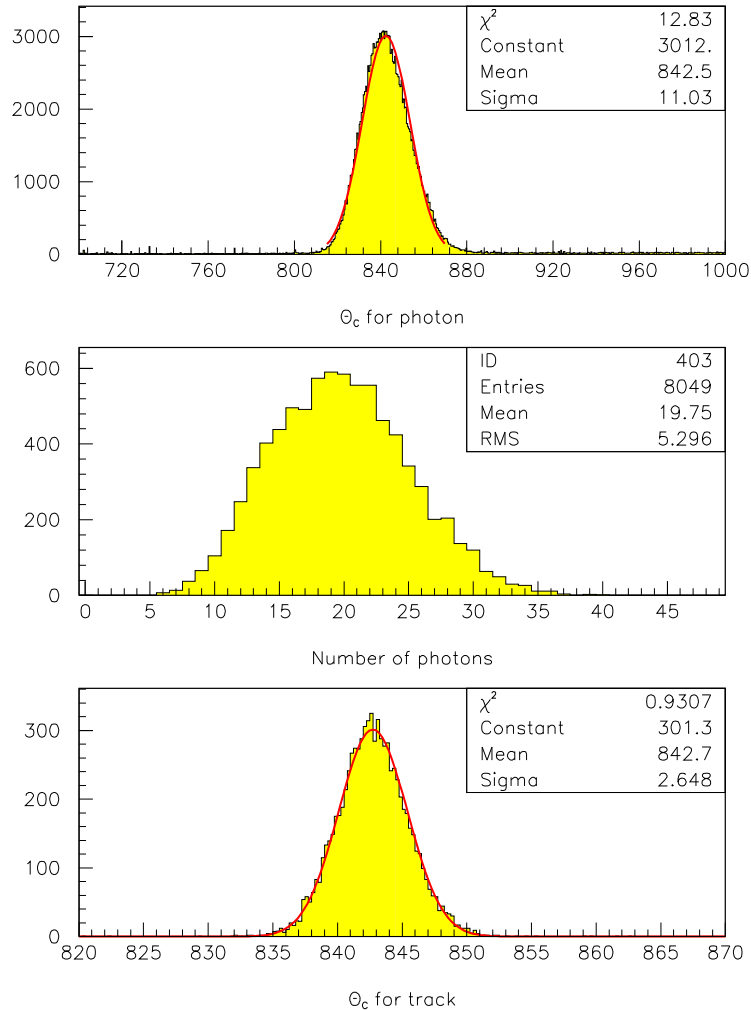

Figure 5: The resolution per photoelectron, number of photoelectrons and Cherenkov angular resolution per track, for an incident track normal to a $45°$ sawtooth radiator.

We see a large average number of photoelectrons. The spread in this distribution is caused in part by the variation in thickness from 7.5 to 12.5 mm. This is reduced in the small tooth design. For these distributions we used a full GEANT simulation including clustering of the pad hits into detected photons, or photoelectrons. This causes a loss in resolution, but the resulting 2.65 mr, is much better than 3.7 mr. The number of photoelectrons, before and after clustering, is shown in Fig. 6. The clustering loss may be ameliorated by better software algorithms. The average resolution per track is shown in Fig. 7 as a function of $\theta$ with the components of the resolution indicated. Some photons are lost due to the finite length of the detector. Some of this loss may be recoverable by mirroring the ends of the detector. The resolution can be improved by making the radiator thinner above $cos(\theta)$ of 0.6, since the emission point error is the dominant contribution in this region.

Tolerances in the manufacture of grooved radiator structures are important. The resolution will worsen if the flat edges of the groove vary by more than ±0.003 mr (rms), or the edges are not parallel to ±0.003 mr (rms). The groove depth can vary as this dimension is not critical. We are working with samples machined by the Center for Optics Manufacturing [6] using material from OPTOVAC [7].

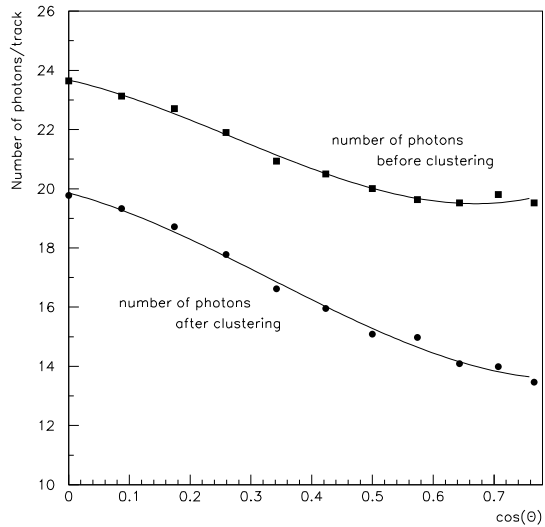

Figure 6: The number of photoelectrons hiting the detector (before clustering) and the number reconstructed by the pattern recognition program (after cluster) as a function of the incident track angle.

## IV. Conclusions

Simulations have shown that a multifacted radiator with 45° teeth gives substantially more photons and better angular resolution per photon than plane crystal radiators. It is also interesting to see what the images look like. In Fig. 8 we show the light pattern for a track normal to the radiator. Recall, that for a flat plane radiator no light exits from radiator surface. The image consists of two intense hyperbolas resulting from light which directly exits the radiator surface, and two lightly populated hyperbolas which result from photons which experience one reflection from the sawtooth surface, either before or after exiting from the surface. There is only ≈10% of the light in these more extended curves.

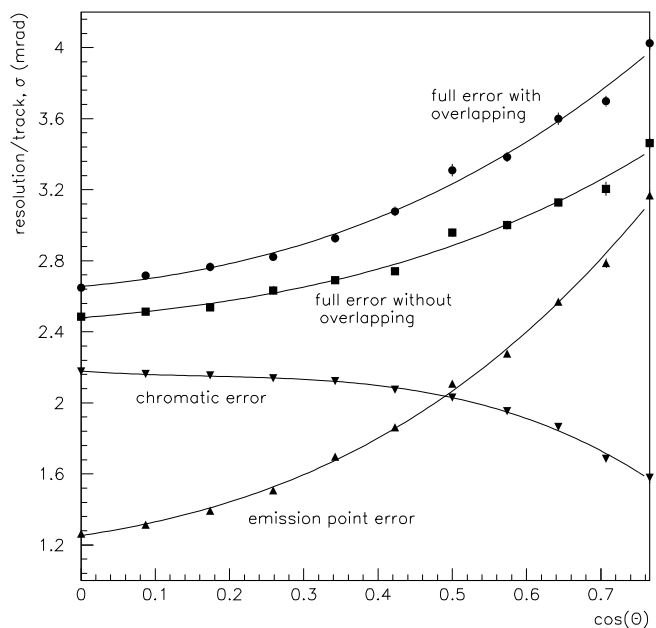

Figure 7: The angular resolution per track as a function of incident track angle. "Without overlapping" shows the error one could obtain with a perfect clustering algorithm, while "with overlapping" shows the effect on the clustering algorithm of having overlapping photoelectrons.

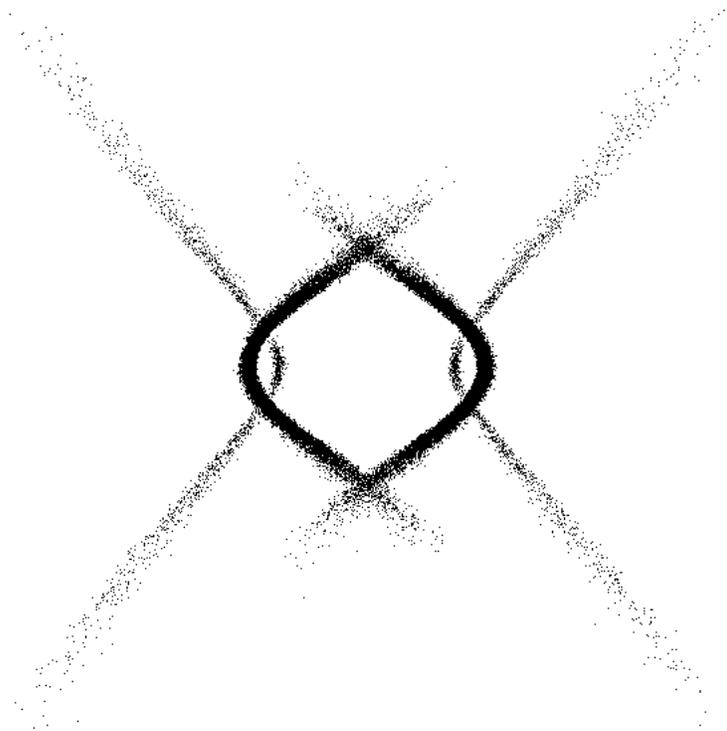

Figure 8: The image pattern for tracks normal to the radiator.